\documentstyle[12pt]{article}
\newcommand{\eq}{\begin{equation}}
\newcommand{\en}{\end{equation}}
\newcommand{\eqa}{\begin{eqnarray}}
\newcommand{\ena}{\end{eqnarray}}
\newcommand{\eqs}{\begin{displaymath}}
\newcommand{\ens}{\end{displaymath}}
\newcommand{\eqas}{\begin{eqnarray*}}
\newcommand{\enas}{\end{eqnarray*}}

\advance\oddsidemargin by -\textwidth
\advance\oddsidemargin by -1in
\evensidemargin=\oddsidemargin
\begin{document}

$\mbox{ }$
\vspace{-3cm}
\begin{flushright}
\begin{tabular}{l}
{\bf KEK-TH-570 }\\
{\bf KEK preprint 98 }\\

\end{tabular}
\end{flushright}
 
\baselineskip18pt
\vspace{1cm}
\begin{center}
\Large
{\baselineskip26pt \bf 
$p$-branes from $(p-2)$-branes \\
in the Bosonic String Theory
}
\end{center}
\vspace{1cm}
\begin{center}
\large
{ Nobuyuki Ishibashi}
\end{center}
\normalsize
\begin{center}
{\it KEK Theory Group, Tsukuba, Ibaraki 305, Japan}
\end{center}
\vspace{2cm}
\begin{center}
\normalsize
ABSTRACT
\end{center}
{\rightskip=2pc 
\leftskip=2pc 
\normalsize
We show that Dirichlet $p$-brane can be expressed as a configuration of 
infinitely many Dirichlet $(p-2)$-branes in the bosonic string theory. 
Using this fact, we interpret the massless fields on the $p$-brane 
worldvolume as deformations of the configuration of the $(p-2)$-branes. 
Especially we find that the worldvolume gauge field  
parametrizes part of the group of diffeomorphisms on the worldvolume. 
\vglue 0.6cm}

\newpage
\section{Introduction}
D-branes\cite{pol} play very important roles in the study of string theory 
dynamics. 
Especially, lower dimensional D-branes such as D-strings, D$0$-branes and 
D-instantons are used in such a way that 
various matrix models are constructed by taking them as the fundamental degrees 
of freedom\cite{BFSS}\cite{IKKT}\cite{DVV}. 
Lower dimensional D-branes are useful because it is easy to quantize the 
worldvolume theory. 
In constructing a theory of such lower dimensional branes, 
it is important to check if 
higher dimensional branes exist in the spectrum. 
The theory lacks something if it does not have higher dimensional branes, 
since superstring theory 
possesses of branes with various dimensions as classical configurations. 

In \cite{townsend}\cite{BFSS}, it was pointed out that D$2$-brane can be 
realized as a 
configuration of infinitely many D$0$-branes in Type IIA/M theory. 
The unbroken supersymmetry in this configuration and  
the relation to the light-cone gauge supermembrane Hamiltonian\cite{dhn} 
confirm that 
this configuration really corresponds to D$2$-brane. 
In \cite{IKKT}, 
the authors showed that D-string can be expressed in a similar way 
as a classical configuration 
of infinitely many D-instantons in the Type IIB case. 
Such relationships between D-branes are generalized 
so that Dirichlet $p$-brane can be expressed as a classical configuration of 
infinitely many Dirichlet $(p-2k)$-branes for $k=1,2,\cdots $. 
What we would like to do in this paper 
is to examine these facts in the framework of string perturbation theory. 

The string theory in the presence of infinitely many D-branes corresponds to 
an open string theory with infinitely many Chan-Paton factors. The 
configuration in question can be realized as a background 
in the open string theory. We would like to check if such a background is really 
equivalent to the string theory in the presence of a D-brane with dimension 
higher by two. In this paper we will show such an equivalence for the bosonic 
string. The superstring case will be considered in a separate publication. 

If one regards a Dirichlet $p$-brane as a configuration of Dirichlet $(p-2k)$-
branes, fluctuations of the $p$-brane can be interpreted as 
fluctuations of the configuration of $(p-2k)$-branes. The worldvolume theory of 
a D-brane is expressed by a $U(1)$ gauge field and scalars. 
In the latter part of this paper we will study what these fields correspond 
to if one considers this D-brane as a configuration of lower dimensional branes. 
Using this correspondence, 
we can clarify the relationship between the gauge field on D-brane worldvolume 
and the reparametrization invariance on the worldvolume.

The organization of this paper is as follows. In section 2, we consider 
a configuration of infinitely many D-branes and show that it yields a higher 
dimensional 
D-brane in the bosonic string case. 
In section 3, we study the worldvolume theory of a D-brane regarding it 
as a configuration of lower dimensional D-branes. 
Section 4 is devoted to the discussions.

\section{$p$-branes from $(p-2)$-branes}
In this section we will show that Dirichlet $p$-brane can be expressed as a 
configuration of infinitely many Dirichlet $(p-2)$-branes in the bosonic string 
theory. 
We will do so for $p=1$ for simplicity. It is straightforward to 
generalize the discussions to other $p$'s. 

The configuration of infinitely many D-instantons can be expressed by the 
$\infty \times \infty$ hermitian matrices $M^\mu~(\mu=1,\cdots ,26)$.
\footnote{Here 
we use $M^\mu$ instead of $X^\mu$ in order not to be confused 
with the string coordinates $X^\mu (\sigma )$. We consider 
the string theory with the Euclidean signature because we are going to study 
D-instantons. Accordingly what we mean by D-string in the following is 
just a two 
dimensional wall.}
The one we 
consider is 
\eqa
M^1=P,
\nonumber
\\
M^2=Q,
\label{PQ-1}
\ena
where $P,Q$ satisfy 
\eq
[Q,P]=ki.
\label{com}
\en 
Here $k$ is a real number. 
What we will do is to consider the bosonic string theory  
corresponding to the configuration eq.(\ref{PQ-1}) of D-instantons 
and show that it is equivalent to the bosonic string theory in the presence of 
a D-string. 

\subsection{The Boundary State}
The interaction between bosonic string and D-brane is described by the 
boundary state. Let us first examine what is the boundary state corresponding 
to the above configuration. 
 
A D-instanton at the origin 
corresponds to the boundary state:
\eqs
|B>_{-1}=|X=0>\otimes |B>_{gh}.
\ens
Here we define $|X=f>$ to be the coherent state satisfying 
\eq
X^\mu (\sigma )|X=f>=f^\mu (\sigma )|X=f>~(\mu =1,\cdots ,26),
\label{coh}
\en
and $|B>_{gh}$ is the ghost boundary state satisfying
\eqs
(c_n+\tilde{c}_{-n})|B>_{gh}=(b_n-\tilde{b}_{-n})|B>_{gh}=0.
\ens
A configuration of $N$ 
D-instantons is expressed by $N\times N$ matrices $M^\mu$ and the 
boundary state for such a configuration is
\eqs
|B>=TrPe^{-\frac{i}{2\pi}\int_0^{2\pi}d\sigma P^\mu(\sigma )M_{\mu}}|B>_{-1},
\ens
where $P^\mu(\sigma )=p^\mu +
\sum_n\frac{1}{2}(\alpha_n^\mu +\tilde{\alpha}_{-n}^\mu )e^{-in\sigma }$.

Now let us consider the boundary state corresponding to the configuration 
eq.(\ref{PQ-1}). We should consider the case $N=\infty$. 
The trace of the path ordered product can be  
expressed by using the path integral representation as
\footnote{The technique to treat such a non-abelian background of open string 
theory was studied previously in \cite{dorn}.}
\eq
|B>_1=\int [dpdq]
e^{\frac{i}{k}\int d\sigma p\partial_\sigma q
-\frac{i}{2\pi}\int_0^{2\pi}d\sigma (P^1p+P^2q)}|B>_{-1}.
\label{path}
\en
Notice that the normalization of $|B>_1$ is ambiguous in this representation 
unless we are careful about the path integral measure $\int [dpdq]$. 
Here we will not pay much attention to the 
normalization of the boundary state. 
More complete argument will be given in the next subsection. 

In this form, it is easy to see that this boundary state coincides with the 
boundary state corresponding to a D-string worldsheet with the background 
$U(1)$ gauge field 
$F_{12}=\frac{4\pi}{k}$, up to normalization. Indeed from the identities 
\eqa
0
&=&
\int [dpdq]\frac{\delta}{\delta p(\sigma )}
e^{\frac{i}{k}\int d\sigma p\partial_\sigma q
-\frac{i}{2\pi}\int_0^{2\pi}d\sigma (P^1p+P^2q)}|B>_{-1},
\nonumber
\\
0
&=&
\int [dpdq]\frac{\delta}{\delta q(\sigma )}
e^{\frac{i}{k}\int d\sigma p\partial_\sigma q
-\frac{i}{2\pi}\int_0^{2\pi}d\sigma (P^1p+P^2q)}|B>_{-1},
\nonumber
\ena
one obtains 
\eqa
& &
[P^1(\sigma )-\frac{2\pi}{k}\partial_\sigma X^2(\sigma )]|B>_1=0,
\nonumber
\\
& &
[P^2(\sigma )+\frac{2\pi}{k}\partial_\sigma X^1(\sigma )]|B>_1=0.
\nonumber
\ena
Solving these equations, one can show that $|B>_1$ corresponds to a D-string 
worldsheet with $F_{12}=\frac{4\pi}{k}$. 

We can check this fact also by actually 
performing the path integral in eq.(\ref{path}). 
Since $|B>_{-1}$ can be obtained up to normalization by solving eqs.(\ref{coh}), 
eq.(\ref{path}) becomes 
\eqa
& &
|B>_1
\propto
\int [dpdq]
e^{\frac{1}{k}\sum_nnq_np_{-n}+
\sum_{n>0}(\frac{1}{n}\alpha_{-n}^\mu\tilde{\alpha}_{-n,\mu} 
-i\alpha_{-n}^1p_n-i\tilde{\alpha}_{-n}^1p_{-n}
-i\alpha_{-n}^2q_n-i\tilde{\alpha}_{-n}^2q_{-n}
-\frac{n}{2}q_nq_{-n}-\frac{n}{2}p_np_{-n})}
\nonumber
\\
& &
\hspace{2.5cm}\times
|x^1=p_0,~x^2=q_0,~x^3=\cdots =x^{26}=0>\otimes |B>_{gh},
\nonumber
\ena
where
\eqa
& &
p(\sigma )=\sum_np_ne^{-in\sigma },
\nonumber
\\
& &
q(\sigma )=\sum_nq_ne^{-in\sigma }.
\nonumber
\ena
The integrations to be done are the Gaussian integrations and 
after the usual procedure of the zeta function regularization, we obtain
\eqa
& &
|B>_1\propto (1+\frac{4}{k^2})^{\frac{1}{2}}
e^{\sum_{n>0}(\frac{2}{n(1+\frac{4}{k^2})}
(\alpha_{-n}^1\tilde{\alpha}_{-n}^1
+\alpha_{-n}^2\tilde{\alpha}_{-n}^2
+\frac{2}{k}
(\alpha_{-n}^1\tilde{\alpha}_{-n}^2-\alpha_{-n}^2\tilde{\alpha}_{-n}^1))
-\frac{1}{n}\alpha_{-n}^\mu\tilde{\alpha}_{-n,\mu}})
\nonumber
\\
& &
\hspace{2.5cm}\times
|p^1=p^2=0,~x^3=\cdots =x^{26}=0>\otimes |B>_{gh}.
\label{stB}
\ena
This is exactly the boundary state for the D-string worldsheet. 

\subsection{Equivalence of the Open String Theories}
The arguments in the previous subsection can show that the boundary state 
for the background eq.(\ref{PQ-1}) coincides with the one for D-string  
only up to normalization. The problem is that we need degrees of 
freedom parametrized by a continuous parameter ($p$ or $q$) to express 
the background eq.(\ref{PQ-1}). This makes the normalization of 
the boundary state with $M^\mu =0$ ambiguous. The most obvious way 
to avoid such ambiguities is to look at things in the open string channel. 
In this subsection, we will construct the open string theory corresponding 
to the D-instantons in the background eq.(\ref{PQ-1}) and show that it is 
equivalent to the open string theory corresponding to the D-string with 
$F_{12}=\frac{4\pi}{k}$. 

The bosonic string theory with a D-instanton at the origin 
can be described by an open bosonic string theory with the Dirichlet boundary 
conditions:
\eqs
X^\mu (\sigma =\pi )=X^\mu (\sigma =0)=0~(\mu =1,\cdots ,26).
\ens
When there are $N$ instantons, one should consider an open string theory 
with the Chan-Paton factors $a=1,\cdots ,N$ and a state in the open string 
Hilbert space is labeled by using two Chan-Paton factors as $|~>_{a\bar{b}}$. 
In order to express the background eq.(\ref{PQ-1}), we need 
the Chan-Paton factors parametrized by a continuous parameter.  
Moreover, since a nontrivial commutation relation eq.(\ref{com})  is involved, 
one should consider an open string theory with quantum mechanical degrees 
of freedom sitting on the boundaries. Namely we will start with the following 
action\footnote{In this subsection, ghosts are ignored because they play no 
role.} 
\eq
I=\frac{1}{8\pi}\int d^2\sigma 
(\dot{X}^\mu \dot{X}_\mu -X^{\prime \mu}X^{\prime}_\mu )
+\frac{1}{k}\int d\tau (p_\pi \dot{q}_\pi -p_0\dot{q}_0).
\en
$p_\pi ,~q_\pi$ and $p_0,~q_0$ are the degrees of freedom sitting on the 
ends $\sigma =\pi$ and $\sigma =0$ respectively. 
Canonically quantizing them, 
we obtain the commutation relations 
\eqa
& &
[q_\pi ,p_\pi ]=ki,
\nonumber
\\
& &
[q_0,p_0]=-ki.
\label{pqcom}
\ena
The difference in the orientation of the two boundaries 
$\sigma =\pi$ and $\sigma =0$ causes the difference in the sign of the above 
commutation relations. The basis of the states may be taken  
to be the eigenstates of $q_\pi$ and $q_0$. Then a state in the open string 
Hilbert space is of the form $|~>_X\otimes |q_\pi ,q_0>$, where $|~>_X$ 
is a state in the Fock space of $X$. Hence $q_\pi$ and $q_0$ behave as  
continuous Chan-Paton factors.

For the background eq.(\ref{PQ-1}), we should consider the following 
action:
\eqa
I
&=&
\frac{1}{8\pi}\int d^2\sigma 
(\dot{X}^\mu \dot{X}_\mu -X^{\prime \mu}X^{\prime}_\mu )
\nonumber
\\
& &
-\frac{1}{4\pi}\int_{\sigma =\pi}d\tau (X^{\prime 1}p_\pi +X^{\prime 2}q_\pi 
+V_\pi (p_\pi ,q_\pi ))
\nonumber
\\
& &
+\frac{1}{4\pi}\int_{\sigma =0}d\tau (X^{\prime 1}p_0 +X^{\prime 2}q_0 
+V_0(p_0,q_0))
\nonumber
\\
& &
+\frac{1}{k}\int d\tau (p_\pi \dot{q}_\pi -p_0\dot{q}_0).
\label{actionD}
\ena
with the boundary conditions 
\eq
X^\mu |_{\sigma =0,~\pi }=0. 
\label{bcD}
\en
Here $V_0$ and $V_\pi$ are divergent counterterms. We will see that these 
terms are necessary. 

What we would like to show in this subsection is that this open string theory 
is equivalent to the open string theory corresponding to a D-string worldsheet 
with 
$F_{12}=\frac{4\pi}{k}$. The action for such an open string theory is 
\eqa
I
&=&
\frac{1}{8\pi}\int d^2\sigma 
(\dot{X}^\mu \dot{X}_\mu -X^{\prime \mu}X^{\prime}_\mu )
\nonumber
\\
& &
-\frac{1}{4\pi}\int_{\sigma =\pi}d\tau \dot{X}^1X^2F_{12}
\nonumber
\\
& &
+\frac{1}{4\pi}\int_{\sigma =0}d\tau \dot{X}^1X^2F_{12},
\label{actionN}
\ena
with the boundary conditions 
\eqa
& &
\partial_{\sigma}X^1|_{\sigma =0,\pi}=
\partial_{\sigma}X^2|_{\sigma =0,\pi}=0,
\nonumber
\\
& &
X^i|_{\sigma =0,\pi}=0~(i=3,\cdots ,26).
\label{bcN}
\ena
We will show that the two dimensional field theory with the action  
in eq.(\ref{actionD}) and the boundary condition in eq.(\ref{bcD}) is 
equivalent to the one with the action in eq.(\ref{actionN}) and the 
boundary condition in eq.(\ref{bcN}). 
Since 
the equivalence is trivially seen for the variables $X^i~(i=3,\cdots ,26)$, 
we will concentrate on $X^1$ and $X^2$ in the following. 

Before doing so, we should remark on the treatment of the boundary terms. 
Dealing with the action in eq.(\ref{actionN}), one usually considers that 
the boundary conditions in eq.(\ref{bcN}) are modified because of the boundary 
terms (the second and the third terms in eq.(\ref{actionN})). 
The new boundary conditions are chosen 
so that the equations of motion is the free one. Thus one obtains 
\eq
(\partial_{\sigma}X^1-F_{12}\partial_{\tau}X^2)|_{\sigma =0,\pi}=
(\partial_{\sigma}X^2+F_{12}\partial_{\tau}X^1)|_{\sigma =0,\pi}=0.
\label{bcF}
\en
Alternatively one can use the boundary conditions 
eq.(\ref{bcN}) and take the effects of the boundary terms into account 
as interaction terms in the equations of motion. 
In this case, the contributions of the boundary terms to the equations 
of motion are proportional to $\delta (\sigma ),~\delta (\sigma -\pi )$, 
and we need to be careful about the treatment of these terms. 
In order to avoid subtlety, we should first shift the positions of the boundary 
terms to be $\sigma =\epsilon$ and $\sigma =\pi -\epsilon ,~(\epsilon >0)$ 
and take the limit $\epsilon \rightarrow 0$ later. 
This ensures that the boundary terms contribute to the equations of motion, 
instead of changing the boundary conditions. 
Of course these two approaches give equivalent results. 
(We will comment on this fact later.) 

Now let us analyze the 2D field theory in eq.(\ref{actionD}). Since there are 
quantum mechanical degrees of freedom on the boundaries, only the 
latter approach in the above paragraph is possible. 
Let us provisionally assume that $V_0=V_\pi =0$. 
Defining the new variables
\eqa
& &
Z=X^1+iX^2,
\nonumber
\\
& &
s_\pi =p_\pi +iq_\pi ,
\nonumber
\\
& &
s_0 =p_0 +iq_0,
\ena
for convenience, the equations of motion become
\eqa
& &
-\ddot{Z}+Z^{\prime \prime}
+s_\pi \delta^{\prime}(\sigma -(\pi -\epsilon ))
-s_0 \delta^{\prime}(\sigma -\epsilon )=0,
\nonumber
\\
& &
\frac{i}{k}\dot{s}_\pi +\frac{1}{4\pi }Z^\prime (\sigma =\pi -\epsilon )=0,
\nonumber
\\
& &
\frac{i}{k}\dot{s}_0+\frac{1}{4\pi }Z^\prime (\sigma =\epsilon )=0,
\label{equations}
\ena
and their complex conjugates. The positions of boundaries are shifted 
by $\epsilon$ as was mentioned in the above remarks. 
From the first equation in eqs.(\ref{equations}), one can see that $Z$ has 
discontinuities like 
\eq
Z=
(s_0+\frac{\sigma -\epsilon}{\pi -2\epsilon}(s_\pi -s_0))
\theta (\sigma -\epsilon )\theta (\pi -\epsilon -\sigma )+\cdots ,
\en
at $\sigma =\epsilon $ and $\sigma =\pi -\epsilon$. Such discontinuities 
make the vertex operators $Z^\prime (\sigma =\epsilon )$ and 
$Z^\prime (\sigma =\pi -\epsilon )$ divergent and make the latter two 
equations in eqs.(\ref{equations}) ill-defined.\footnote{We would like to 
thank H. Dorn for pointing out this problem.}

Since we did not encounter such divergences in the previous subsection, there 
should be discrepancy between the vertex operator $P^\mu$ in the closed 
string channel and the vertex operator $X^{\prime\mu}$. Indeed 
$P^\mu$ in the closed string channel corresponds to 
$\lim_{\eta\rightarrow 0_+}X^\mu (\sigma =\epsilon +\eta )$ and 
$\lim_{\eta\rightarrow 0_+}X^\mu (\sigma =\pi -\epsilon -\eta )$ in the 
open string channel and the divergences coming from the discontinuities 
are avoided. Therefore we will modify the action so 
that the vertex operators coincide with 
those in the closed string channel. 
Accordingly the counterterms in eq.(\ref{actionD}) should be 
\eqa
& &
V_\pi =\frac{1}{2}p^2_\pi \delta (0)
+\frac{1}{2}q^2_\pi \delta (0),
\nonumber
\\
& &
V_0=-\frac{1}{2}p^2_0 \delta (0)
-\frac{1}{2}q^2_0 \delta (0).
\ena
With this modification, the second and the third equations in 
eqs.(\ref{equations}) become 
\eqa
& &
\frac{i}{k}\dot{s}_\pi 
+\frac{1}{4\pi }\lim_{\eta \rightarrow 0_+}Z^\prime 
(\sigma =\pi -\epsilon -\eta )=0,
\nonumber
\\
& &
\frac{i}{k}\dot{s}_0
+\frac{1}{4\pi }\lim_{\eta\rightarrow 0_+}Z^\prime (\sigma =\epsilon +\eta )=0.
\ena

The boundary conditions $Z|_{\sigma =0,\pi}=0$ 
implies that $Z$ can be expanded as 
\eq
Z(\tau ,\sigma )=2\sqrt{2}\sum_{n>0}z_n(\tau )\sin n\sigma .
\label{expz}
\en
In terms of $z_n$, the equations of motion are 
\eqa
& &
\ddot{z}_n+n^2z_n
+\frac{n}{\sqrt{2}\pi}(s_\pi \cos n(\pi -\epsilon )-s_0\cos n\epsilon )=0,
\nonumber
\\
& &
\dot{s}_\pi =\frac{ik}{\sqrt{2}\pi}\sum_{n>0}nz_n\cos n(\pi -\epsilon -\eta ),
\nonumber
\\
& &
\dot{s}_0 =\frac{ik}{\sqrt{2}\pi}\sum_{n>0}nz_n\cos n(\epsilon +\eta ),
\label{eqmz}
\ena
and the canonical commutation relations are
\eqa
\mbox{[}z_n,\dot{z}_m^\dagger\mbox{]}
&=&
2i\delta_{n,m},
\nonumber
\\
\mbox{[}z_n^\dagger ,\dot{z}_m\mbox{]}
&=&
2i\delta_{n,m},
\nonumber
\\
& &
0,~\mbox{otherwise.}
\label{comz}
\ena

On the other hand, the action in eq.(\ref{actionD}) can be treated 
in the usual manner\cite{callan}. The boundary term modifies the 
boundary condition as in eq.(\ref{bcF}), which is 
\eq
(Z^\prime +iF\dot{Z})|_{\sigma =0,\pi}=0~(F\equiv F_{12}),
\en
in terms of $Z=X^1+iX^2$. Then $Z$ can be expanded as 
\eq
Z(\tau ,\sigma )
=
2\sqrt{2}\{
\frac{x_++p_-\mbox{[}\tau -iF(\sigma -\frac{\pi}{2})\mbox{]}}{\sqrt{1+F^2}}
+i\sum_{n=1}^{\infty}\mbox{[}
\frac{a_n(\tau )}{\sqrt{n}}\cos (n\sigma +\gamma )
-\frac{b_n^\dagger (\tau )}{\sqrt{n}}\cos (-n\sigma +\gamma )
\mbox{]}\},
\label{expab}
\en
where $\tan \gamma =F$. In terms of the coefficients of this expansion, 
the equations of motion become 
\eqa
& &
(\partial_\tau +in)a_n=0,
\nonumber
\\
& &
(\partial_\tau -in)b_n^\dagger =0.
\label{eqmab}
\ena
The canonical commutation relations are
\eqa
\mbox{[}a_n,a_m^\dagger \mbox{]}
&=&
\delta_{n,m},
\nonumber
\\
\mbox{[}b_n,b_m^\dagger \mbox{]}
&=&
\delta_{n,m},
\nonumber
\\
\mbox{[}x_+,p_+\mbox{]}=\mbox{[}x_-,p_-\mbox{]}
&=&
i,
\nonumber
\\
& &
0~\mbox{otherwise,}
\label{comab}
\ena
where $x_-\equiv (x_+)^\dagger ,~p_+\equiv (p_-)^\dagger$. 

Now we will show that these two systems are equivalent if $kF=4\pi$. 
Actually we can show that there exist a transformation from the variables 
$(z_n,~\dot{z}_n,~s_\pi ,~s_0)$ to the variables 
$(a_n,~b_n^\dagger ,~x_+,~p_-)$ which turns 
eqs.(\ref{pqcom}), (\ref{expz}), (\ref{eqmz}), (\ref{comz}) into 
eqs.(\ref{expab}), (\ref{eqmab}), (\ref{comab}). 
In order to do so, 
one should notice the following fact. The boundary terms are treated 
quite differently in these two systems. 
In the former case, the positions of the boundaries are shifted a 
bit and the boundary terms contribute to the equations of motion. 
In the latter case, the positions of the boundaries are not shifted and 
the boundary conditions are changed. Therefore, the variable $Z$ in 
eq.(\ref{expz}) and eq.(\ref{expab}) can be compared only in the region 
$\epsilon \leq \sigma \leq \pi -\epsilon$. As a function in such a region, 
$Z$ in eq.(\ref{expab}) can be expanded by $\sin n\sigma$ 
\footnote{If one expands 
eq.(\ref{expab}) in the same spirit in terms of $\cos n\sigma$, one 
can show the equivalence of the two approaches to deal with the boundary 
terms in eq.(\ref{actionN}). }
and we obtain the 
following invertible transformation from the variables 
$(z_n,~\dot{z}_n,~s_\pi ,~s_0)$ to the variables 
$(a_n,~b_n^\dagger ,~x_+,~p_-)$ as long as $F\neq 0,~k\neq 0$:
\eqa
a_n
&=&
\frac{1}{2\sqrt{n(1+F^2)}}
(\dot{z}_n-inz_n
-\frac{2}{F\pi}\sum_{m>0}\frac{m}{m^2-n^2}(\dot{z}_m-inz_m)(1-(-1)^{n+m})
\nonumber
\\
& &
-\frac{i}{\sqrt{2}\pi}((-1)^ns_\pi -s_0)),
\nonumber
\\
b_n^\dagger
&=&
\frac{1}{2\sqrt{n(1+F^2)}}
(\dot{z}_n+inz_n
+\frac{2}{F\pi}\sum_{m>0}\frac{m}{m^2-n^2}(\dot{z}_m+inz_m)(1-(-1)^{n+m})
\nonumber
\\
& &
+\frac{i}{\sqrt{2}\pi}((-1)^ns_\pi -s_0)),
\nonumber
\\
x_+
&=&
\frac{1}{\pi \sqrt{1+F^2}}\sum_{m>0}z_m\frac{1-(-1)^m}{m}
+\frac{Fi}{2\sqrt{1+F^2}}\sum_{m>0}\dot{z}_m\frac{1+(-1)^m}{m}
\nonumber
\\
& &
+\frac{F^2}{4\sqrt{2(1+F^2)}}(s_\pi +s_0),
\nonumber
\\
p_-
&=&
\frac{1}{\pi \sqrt{1+F^2}}\sum_{m>0}\dot{z}_m\frac{1-(-1)^m}{m}
-\frac{Fi}{2\pi\sqrt{2(1+F^2)}}(s_\pi -s_0).
\label{DNtrans}
\ena
After tedious but straightforward calculations, one can show that 
eqs.(\ref{pqcom}), (\ref{eqmz}), (\ref{comz}) are transformed into 
eqs.(\ref{eqmab}), (\ref{comab}) by this transformation. 

We conclude this section by the following remark. 
What we have shown in this section is that the background of the form 
eq.(\ref{PQ-1}) turns the Dirichlet boundary conditions of $X^1,~X^2$ 
into the Neumann boundary conditions with the background gauge field 
$F_{12}=\frac{4\pi}{k}$. 
Using this background, it is straightforward to show that Dirichlet 
$p$-brane with 
a background gauge field can be obtained as a configuration of infinitely 
many Dirichlet $(p-2)$-branes. By repeating this procedure,  
Dirichlet $p$-brane can be obtained as a configuration 
of infinitely many Dirichlet $(p-2n)$-branes for $(n=1,2,\cdots )$.

\section{Worldvolume Theory}
Thus far, we consider a straight Dirichlet $p$-brane at rest and show that 
it can be realized as a configuration of infinitely many Dirichlet 
$(p-2)$-branes, by proving the 
equivalence of the string theories in the two backgrounds. Since the modes 
to deform these backgrounds are included in the open string spectrum, 
it is possible to regard a more general configuration of $p$-brane 
worldsheet 
as a configuration of $(p-2)$-branes. Hence everything about $p$-brane can be
reinterpreted by regarding it as a configuration of $(p-2)$-branes. 
Namely there are two ways of looking at the same thing, 
either as a $p$-brane or as a configuration of $(p-2)$-branes. 
From now on, let us call these two points of view $p$-brane picture and 
$(p-2)$-brane picture respectively. 
In this section we will examine what the fields on the worldvolume of 
Dirichlet $p$-brane correspond to in the $(p-2)$-brane picture. 
We will do so for $p=1$ for simplicity but the results here can be 
easily generalized to other $p$'s. 

The worldsheet field content of D-string consists of a worldsheet vector 
$A_\alpha ~(\alpha =1,2)$ and scalars $\phi^i~(i=3,\cdots ,26)$. 
$\phi^i$ can be interpreted as oscillations in the position of the D-string. 
Since the open string vertex operator corresponding to $\phi^i$ is 
$\partial_\sigma X^i\phi_i(X)$, 
the boundary state corresponding to the D-string worldsheet with the shape 
specified by the equations $X^i=\phi^i(X^\alpha )$ is
\eqs
|B>=e^{-\frac{i}{2\pi}\int_0^{2\pi}d\sigma P_i(\sigma )\phi^i(X)}|B>_1.
\ens
For our purpose, it is convenient to rewrite $|B>$ by using eq.(\ref{path}) as 
\eq
|B>=\int [dpdq]
\exp \mbox{[}\frac{i}{k}\int d\sigma p\partial_\sigma q
-\frac{i}{2\pi}\int_0^{2\pi}d\sigma (P^1p+P^2q+P_i\phi^i(p,q))\mbox{]}
|B>_{-1},
\label{Bphi}
\en
where $\phi (p,q)\equiv \phi (X^1=p,~X^2=q)$. 

The part  
$e^{-\frac{i}{2\pi}\int_0^{2\pi}d\sigma (P^1p+P^2q+P_i\phi^i(p,q))}|X=0>$ 
in the integrand is the coherent state 
$|X^1=p(\sigma ),~X^2=q(\sigma ),~X^i=\phi^i(p(\sigma ),~q(\sigma ))>$. 
Therefore the fact that 
the fundamental string is emitted from the worldsheet $X^i=\phi^i(X^\alpha )$ 
is obvious in this form. Moreover since it is in the form of the  
$|B>_{-1}$ deformed by vertex operators, 
it is easy to see how scalars 
$\phi^i$ 
can be interpreted as deformations of the configuration eq.(\ref{PQ-1}) of 
D-instantons. It is obvious that $|B>$ corresponds to a background 
\eqa
& &
M^1=P,
\nonumber
\\
& &
M^2=Q,
\nonumber
\\
& &
M^i=\phi^i(P,Q),
\label{PQ-1phi}
\ena
with $[Q,P]=ki$. Here the operators are assumed to be Weyl ordered in the 
expression $\phi^i(P,Q)$. 
Therefore the scalars $\phi^i$ can be directly interpreted as 
a deformation of the background eq.(\ref{PQ-1}). 

However $\phi^i$'s do not exhaust all the possible deformations either in 
the D-string picture or the D-instanton picture. 
In the D-instanton picture, the existence of the vertex operator 
$\partial_\sigma X^\mu\phi_\mu (X)~(\mu =1,\cdots ,26)$ implies that the more 
general deformation to be considered is 
\eq
M^\mu =\phi^\mu (P,Q).
\label{PQ-1phimu}
\en
Here $P,Q$ play the role of the coordinates on of the 
worldsheet 
of D-string. In the background in eq.(\ref{PQ-1phi}), $X^1$ and $X^2$ are 
taken to be such coordinates. The background in eq.(\ref{PQ-1phimu}) corresponds 
to a more general parametrization. 
The boundary state corresponding to such a background is 
\eq
|B>=\int [dy]
\exp\mbox{[}\frac{i}{k}\int d\sigma y^1\partial_\sigma y^2
-\frac{i}{2\pi}\int_0^{2\pi}d\sigma P_\mu \phi^\mu (y)\mbox{]}
|B>_{-1},
\label{Bphimu}
\en
where we have switched the notation so that the variables $p,q$ 
in eq.(\ref{Bphi}) correspond to $y^1,y^2$. Accordingly $y^\alpha$'s play the 
role of the coordinates on the worldsheet. 

On the other hand, in the D-string picture, we have 
$A_\alpha$ in addition to $\phi^i$. Since the vertex operator for $A_\alpha$ 
is $\partial_\tau X^\alpha A_\alpha (X)$, the deformation in 
this direction modifies the boundary state in eq.(\ref{Bphimu}) as 
\eq
|B>=\int [dy]
\exp\mbox{[}\frac{i}{4\pi}\int d\sigma A_\alpha (y)\partial_\sigma y^\alpha
-\frac{i}{2\pi}\int_0^{2\pi}d\sigma P_\mu \phi^\mu (y)\mbox{]}|B>_{-1},
\label{BphimuA}
\en
which coincides with eq.(\ref{Bphimu}) when one has $F_{12}=4\pi /k$. 

The problem is what $A_\alpha$ is in the D-instanton picture. 
The vertex operator $\partial_\tau X^\alpha A_\alpha (X)$ in the 
D-string picture becomes $A_\alpha (y)\partial_\sigma y^\alpha$, 
which is not a vertex operator in the D-instanton picture. 
Since the background $A_\alpha$ does not correspond to any vertex operator 
in the D-instanton picture, it appears that $A_\alpha$ cannot be interpreted as 
a deformation of the configuration in eq.(\ref{PQ-1}). 
However one can derive the following identities:
\eqa
0
&=&
\int \mbox{[}dy\mbox{]}\frac{\delta}{\delta y^\alpha}
e^{\frac{i}{4\pi}\int d\sigma A_\alpha (y)\partial_\sigma y^\alpha
-\frac{i}{2\pi}\int_0^{2\pi}d\sigma P_\mu \phi^\mu (y)}|B>_{-1}
\nonumber
\\
&=&
\int \mbox{[}dy\mbox{]}
\mbox{[}\frac{i}{4\pi}F_{\alpha \beta}(y)\partial_\sigma y^\beta
-\frac{i}{2\pi}P_\mu \partial_\alpha\phi^\mu\mbox{]}
e^{\frac{i}{4\pi}\int d\sigma A_\alpha (y)\partial_\sigma y^\alpha
-\frac{i}{2\pi}\int_0^{2\pi}d\sigma P_\mu \phi^\mu (y)}|B>_{-1}.
\nonumber
\ena
Hence the operator 
$\int d\sigma \partial_\tau X^\alpha \delta A_\alpha (X)$ which corresponds to 
variation $\delta A$ in the D-string picture is equivalent to 
\eq
2\int d\sigma 
\delta A_\alpha (F^{-1})^{\alpha\beta}\partial_\beta\phi^\mu (y)P_\mu ,
\label{delA}
\en
when operating on $|B>$. Here we assume that the background 
$F_{\alpha \beta}(y)$ is invertible as a $2\times 2$ matrix for all $y$. 

Therefore, at least perturbatively, variations of the gauge field $A_\alpha$ 
can be interpreted 
as variations of $\phi^\mu$ as long as  $F_{\alpha \beta}(y)$ is invertible. 
This condition for $F_{\alpha \beta}(y)$ has the following meaning. 
Let us take a look at the expression of the boundary state in 
eq.(\ref{BphimuA}). 
Being a path integral over the variables $y^\alpha (\sigma )$, one can evaluate 
the right-hand-side of eq.(\ref{BphimuA}) by canonically quantizing 
$y^\alpha (\sigma )$. In doing so, the symplectic form to be used can be read 
off to be $F_{\alpha\beta}$. Therefore the quantum mechanics of 
$y^\alpha (\sigma )$ is well-defined if $F_{\alpha\beta}$ is invertible. 
This condition is satisfied, for example, by the background in eq.(\ref{PQ-1}). 

Then what kind of deformation of $\phi^\mu$ does $A_\alpha$ correspond to? 
It is easy to see from eq.(\ref{delA}) that the gauge field is related to the 
reparametrization of the worldsheet and the variations $\delta A_\alpha$ are 
equivalent to the variations 
\eq
\delta y^\alpha =(F^{-1})^{\alpha\beta}\delta A_\beta ,
\label{deltay}
\en
of the coordinates $y^\alpha$. This is consistent with the results in 
\cite{miao}, when $F_{\alpha\beta}$ is constant. 

Conversely, we can say that the gauge field 
should be transformed as 
\eq
\delta A_\alpha =F_{\alpha\beta}\delta y^\beta ,
\label{new}
\en
under the reparametrization $y^\alpha \rightarrow y^\alpha +\delta y^\alpha$ 
in the D-instanton picture. 
Eq.(\ref{new}) was considered for the global transformation in \cite{schwarz}. 
This transformation law can be rewritten as 
\eq
\delta A_\alpha =-\partial_\alpha\delta y^\beta A_\beta
-\delta y^\beta\partial_\beta A_\alpha 
+\partial_\alpha (\delta y^\beta A_\beta ),
\en
and the first two terms coincide with the usual variation of $A_\alpha$ 
under $y^\alpha \rightarrow y^\alpha +\delta y^\alpha$. Therefore 
eq.(\ref{new}) gives the usual transformation law for gauge invariant 
quantities. 

Now let us summarize. If one regards D-string as a configuration of infinitely 
many D-instantons, the deformation of such a configuration is parametrized by 
$\phi^\mu (y)~(\mu =1,\cdots ,26)$ in eq(\ref{PQ-1phimu}). 
If there existed invariance under diffeomorphism  
$y^\alpha \rightarrow y^\alpha +\delta y^\alpha$, the space of deformations 
would be 
\eq
\begin{array}{c}
\underline{\mbox{space~of~}\phi^\mu (y)}
\\
Diff
\end{array}
,
\en 
where $Diff$ is the group of diffeomorphisms. 
However in order to regard D-string as a configuration of D-instantons
 (i.e.  in order for eq.(\ref{BphimuA}) to be well-defined), 
one should have $F_{\alpha\beta}(y)$ invertible for all $y$, which breaks the 
reparametrization invariance. Therefore the space of deformations become 
\eq
\begin{array}{c}
\underline{\mbox{space~of~}\phi^\mu (y)}
\\
Diff_{F}
\end{array}
,
\label{instp}
\en
where $Diff_{F}$ denotes the group of diffeomorphisms which preserve 
$F_{\alpha\beta}(y)$. If one takes $F_{\alpha\beta}$ as the 
volume element on the worldsheet, $Diff_{F}$ corresponds to the group of 
area preserving diffeomorphisms. 

On the other hand, the deformations of a D-string worldsheet are parametrized by 
$A_\alpha$ and $\phi^i$. The space of deformations is 
\eq
(\mbox{space~of~}\phi^i(y))\otimes 
\left( \begin{array}{c}
\underline{\mbox{space~of~}A_\alpha (y)}
\\
{\cal G}
\end{array}
\right),
\label{strp}
\en
where ${\cal G}$ denotes the group of gauge transformations. 
Eq.(\ref{instp}) is equivalent to eq.(\ref{strp}). 
Indeed eq.(\ref{instp}) can be rewritten as
\eq
\left( \begin{array}{c}
\underline{\mbox{space~of~}\phi^\mu (y)}
\\
Diff
\end{array}
\right)
\otimes 
\left( \begin{array}{c}
Diff
\\
\overline{Diff_{F}}
\end{array}
\right).
\en
The first factor can be shown to be equivalent to the first factor of 
eq.(\ref{strp}) by taking the static gauge $y^1=\phi^1,~y^2=\phi^2$. 
The second factor is transformed to the second factor of eq.(\ref{strp}) 
via the relation eq.(\ref{deltay}). Indeed the reparametrization which 
preserves $F_{\alpha\beta}(y)$ gives a variation $\delta A_\alpha$ which 
does not change $F_{\alpha\beta}(y)$ i.e. a gauge variation. 
Thus the D-string picture and the D-instanton picture are equivalent. 
These two pictures are related to each other by the nonlinear 
field redefinition in eq.(\ref{deltay}). 

We conclude this section by the following remarks. 
What we have done in this section can be done for any Dirichlet $p$-brane. 
As was mentioned at the end of the previous section, such 
a brane can be considered as a configuration of infinitely many Dirichlet 
$(p-2)$-branes. The worldvolume field content of Dirichlet $p$-brane 
consists of $A_\alpha ~(\alpha =1,\cdots ,p+1)$ and $\phi^i~(i=p+2,\cdots ,26)$. 
$A_\alpha ~(\alpha =1,\cdots ,p-1)$ and $\phi^i~(i=p+2,\cdots ,26)$ are 
common to the $p$-brane and the $(p-2)$-branes. 
Via the relation eq.(\ref{deltay}), 
$A_\alpha ~(\alpha =p,p+1)$ 
on the $p$-brane worldvolume corresponds to part of the group of diffeomorphisms 
in the $p$-th and the $(p+1)$-th directions in the $(p-2)$-brane picture. 

\section{Discussions}
In this paper we have proved that Dirichlet $p$-brane can be considered as a 
configuration of infinitely many Dirichlet $p$-branes. What was essential in 
the proof is the fact that the system eq.(\ref{actionD}) with the 
Dirichlet boundary condition for $X^1,~X^2$ is equivalent to the system 
eq.(\ref{actionN}) with the Neumann boundary condition for  $X^1,~X^2$. 
This correspondence between Dirichlet and Neumann conditions is 
reminiscent of T-duality\cite{pol}\cite{T}. 
However there are several important 
differences. Firstly the correspondence discussed in this paper is between 
the open string with the Dirichlet condition and the open string with 
the Neumann condition with {\it infinitely many Chan-Paton factors}. 
Secondly the variables $X^1,~X^2$ are transformed to themselves in the 
transformation eq.(\ref{DNtrans}). 

Since the open string theories are equivalent, everything about $p$-brane 
can be reinterpreted by regarding it as a configuration of $(p-2)$-branes. 
Particularly when $p$ is odd, Dirichlet $p$-brane can be recognized as 
a configuration of D-instantons. The D-instanton picture is convenient 
when one considers the worldvolume theory of the $p$-brane. The fields 
to be considered are $\phi^\mu$ and the spacetime Lorentz invariance is 
manifest. 
Moreover the locality of the interaction between D-branes and fundamental 
string is manifest in the form of the boundary state eq.(\ref{BphimuA}). 
This form of the boundary state is useful in analysing such interactions. 
The gauge field on the worldvolume of the $p$-brane is 
related to the group of diffeomorphisms on the worldvolume via the 
relation eq.(\ref{deltay}). This relation was derived when there exists 
one  $p$-brane. If there are $N$ $p$-branes, the gauge field becomes 
a $U(N)$ gauge field. It is intriguing to consider the generalization of 
eq.(\ref{deltay}) for this case. 
It seems that the $U(N)$ gauge field should correspond to 
the diffeomorphism group enlarged to include the statistics group 
on $N$ $p$-branes. 

It is straightforward to supersymmetrize all the results in this paper. 
In the supersymmetrized version, we should generalize the notion of 
Chan-Paton factor a little bit. The relation eq.(\ref{deltay}) 
also holds in the superstring case. We will discuss this in a separate 
publication.

\section*{Acknowledgements}
We would like to thank H. Dorn, K. Hamada, H. Kawai, Y. Kazama, H. Kunitomo, 
K. Murakami, M. Natsuume and  T. Oota for useful discussions and comments. 
This work was supported by the Grant-in-Aid for Scientific Research from the 
Ministry of Education, Science and Culture of Japan. 

\newpage
%

\begin{thebibliography}{99}
\bibitem{pol}
J. Dai, R. G. Leigh and J. Polchinski, Mod. Phys. Lett. {\bf A4} (1989) 2073;\\
J. Polchinski, hep-th/9510017, Phys. Rev. Lett. {\bf 75} (1995) 4724.

\bibitem{BFSS}
T. Banks, W. Fischler, S. H. Shenker and L. Susskind, hep-th/9610043, 
Phys. Rev. {\bf D55} 
(1997) 5112.

\bibitem{IKKT}
N. Ishibashi, H. Kawai, Y. Kitazawa and A. Tsuchiya, hep-th/9612115, 
Nucl. Phys. {\bf B498} (1997) 467.

\bibitem{DVV}
R. Dijkgraaf, E. Verlinde and H. Verlinde, hep-th/9703030, 
Nucl. Phys. {\bf B500} (1997) 43. 

\bibitem{townsend}
P. K. Townsend, hep-th/9512062, 
Phys. Lett. {\bf B373} (1996) 68.

\bibitem{dhn}
B. de Wit, J. Hoppe and H. Nicolai, Nucl. Phys. {\bf B305}[FS23] (1988) 545. 

\bibitem{dorn}
H. Dorn and H.-J. Otto, hep-th/9603186, 
Phys. Lett. {\bf B381} (1996) 81; 
hep-th/9702018, 
Nucl. Phys. B(Proc. Suppl.) {\bf 56B} (1997) 30;\\
H. Dorn, hep-th/9612120, Nucl. Phys. {\bf B494} (1997) 105.

\bibitem{callan}
A. Abouelsaood, C. G. Callan, C. R. Nappi and S. A. Yost, Nucl. Phys. 
{\bf B280}[FS18] (1987) 599;\\
V. V. Nesterenko, Int. Journ. Mod. Phys. {\bf A4} (1989) 2627. 

\bibitem{miao}
M. Li, hep-th/9612222, 
Nucl. Phys. {\bf B499} (1997) 149. 

\bibitem{schwarz}
M. Aganagic, C. Popescu and J. H. Schwarz, hep-th/9612080, 
Nucl. Phys. {\bf B495} (1997) 99. 

\bibitem{T}
E. Alvarez, J. L. F. Barbon and J. Borlaf, hep-th/9603089, 
Nucl. Phys. {\bf B479} (1996) 218;\\
H. Dorn and H. J. Otto, hep-th/9603186
Phys. Lett. {\bf B381} (1996) 81;\\
S. Forste, A. A. Kehagias and S. Schwager, hep-th/9604013, 
Nucl. Phys. {\bf B478} (1996) 141, 
hep-th/9610062, hep-th/9611060;\\
J. Borlaf and Y. Lozano, hep-th/9607051, 
Nucl. Phys. {\bf B480} (1996) 239;\\
Y. Lozano, hep-th/9610024, 
Mod. Phys. Lett. {\bf A11} (1996) 2893.

\end{thebibliography}
\end{document}